\documentclass[leqno]{article}%
\usepackage{amsfonts,amsmath,amsthm,amssymb,times}
\usepackage{color}
\usepackage{graphicx,subfigure,color,float,enumerate}
\pdfoutput=1
\newtheorem{theorem}{THEOREM}[section]

\newtheorem{definition}{DEFINITION}[section]
\newtheorem{corollary}{COROLLARY}[section]

\newtheorem{proposition}{PROPOSITION}[section]

\numberwithin{equation}{section}

\title{Nonparametric Confidence Regions for Veronese-Whitney Means and Antimeans on Planar Kendall Shape Spaces}
\author{Yunfan Wang and Vic Patrangenaru}
\date{October 10, 2018}

\begin{document}

\maketitle

\section{Introduction} \label{s:1}

To date, Object Data Analysis (ODA) is the most inclusive type of data analysis, as far as sample metric spaces are concerned. Early examples of object spaces were spaces of directions (see Watson(1983) \cite{Wa:1983}), direct similarity shape spaces (see Kendall(1984)\cite{Ke:1984}), axial spaces (see Beran and Fisher(1998)\cite{BeFi:1998}, Fisher et al.(1996) \cite{FiHaJiWo:1996}),  Stiefel manifolds (see Hendriks and Landsman(1998)\cite{HeLa:1998}).
In the infinite dimensional case, ODA leads to a nonlinear extension of functional data analysis (see Patrangenaru and Ellingson (2015)\cite{PaEl:2015}).

Fr\'echet (1948)\cite{Fr:48} noticed that for higher complexity data, such as the shape of a random contour, numbers or vectors do not provide a meaningful representation. To investigate these kind of data he introduced the notion of {\em elements}, which are nowadays called objects; as an example he mentioned  that ``the shape of an egg randomly taken from a basket of eggs" may be viewed as a random object. Fr\'echet's visionary concepts, were nevertheless hard to handle computationally during his time. It took many decades, until such data became the bread and butter of modern data analysis. Nowadays, various types of shapes of configurations extracted from digital images are represented as points on projective shape spaces (see Mardia and Patrangenaru (2005)\cite{MaPa:2005}, Patrangenaru et al.(2010)\cite{PaLiSu:2010}), on affine shape spaces(see Patrangenaru and Mardia(2003)\cite{PaMa:2003}, Sugathadasa(2006) \cite{Su:2006}), or on Kendall shape spaces (see Kendall(1984) \cite{Ke:1984}, Dryden and Mardia(2016)\cite{DrMa:2016}). To analyze the mean and variance of the random object $X$ on a smooth object space $\mathcal M$ with a metric $\rho$, Fr\'echet defined what we call now the {\em  Fr\'echet function} given by
\begin{equation}\label{eq:frechet-f1}
\mathcal F(p) =  \mathbb E( \rho^2 (p,x)),
\end{equation}
and if $(\mathcal M,\rho)$ is complete, the minimizers of the Fr\'echet function form the {\em Fr\'echet mean set}.
In general, if $\rho = \rho_g$ is the geodesic distance associated with a Riemannian structure $g$ on a manifold $\mathcal M,$ there are no necessary and sufficient conditions for the existence of a unique minimizer of $\mathcal F$ in \eqref{eq:frechet-f1} (see eg Patrangenaru and Ellingson (2015)\cite{PaEl:2015}, ch.4), therefore, with the possible exception of complete flat Riemannian manifolds, it is advisable to consider only the case when $\rho$ is the ``chord" distance on $\mathcal M$ induced  by the Euclidean distance in $\mathbb{R}^{N}$ via an embedding $j: \mathcal{M} \rightarrow \mathbb{R}^{N},$ and the Fr\'echet function becomes
\begin{equation}\label{eq:frechet-f2}
\mathcal{F}(p)= \int_{\mathcal{M}}  \| j(x) - j(p) \|^{2}_{0} Q(dx),
\end{equation}
where $Q = P_X$ is the probability measure on $\mathcal M,$ associated with $X.$
Also, given $X_1, \dots, X_n$ i.i.d.r.o.'s from $Q$, their {\em extrinsic sample mean (set)} is the extrinsic mean (set) of the empirical distribution $\hat Q_n = {\frac{1}{n}} \sum_{i=1}^n \delta_{X_i}$ (see eg Patrangenaru and Ellingson(2015)\cite{PaEl:2015}, chapter 4).\\
In this paper we will assume in addition that $(\mathcal{M}, \rho) $ is a compact metric space, therefore the Fr\'echet function is bounded, and its extreme values are attained two set of points on $\mathcal M.$  It makes sense to also consider as {\bf location parameter} for $X,$  the {\em extrinsic antimean set}, set of maximizers of the Fr\'echet function in \eqref{eq:frechet-f2} (see eg Patrangenaru, Guo and Yao (2016)\cite{PG16}). In case the extrinsic antimean set has one point only, that point is called {\bf extrinsic antimean} of $X,$ and is labeled $\alpha\mu_{j,E}(Q)$, or simply $\alpha\mu_E,$ when $j$ and $Q$ are known.

In this paper after a brief revision of Veronese-Whitney means (VW means) in Section \ref{s:2}, which are extrinsic means on real and complex projective spaces, relative to the Veronese-Whitney embeddings, we give two examples of sample VW means computations on Kendall shape spaces. In Section \ref{s:3} we derive large sample and pivotal nonparametric bootstrap confidence regions for VW antimeans, using VW anticovariance matrices, and their sample counterparts.

\section{VW antimeans on $\mathbb C P ^{k-2}$} \label{s:2}

Planar direct similarity shapes of $k$-ads ( set of $k$ labeled points at least two of which are distinct) in the Euclidean space, were introduced by D. G. Kendall (1984)\cite{Ke:1984}, who showed that in the 2D case, these shapes can be represented as points on a complex projective space $\mathbb CP^{k-2}.$ A standard shape analysis method, due to Kent(1992)\cite{Ke:1992}, consists in using the so called Veronese-Whitney (VW) embedding of $\mathbb CP^{k-2}$ in the space of $(k-1)\times(k-1)$ self adjoint complex matrices, to represent shape data in an Euclidean space. This VW embedding $j$: $\mathbb C P^{k-2} \to S(k-1, \mathbb C)$, where $S(k-1,\mathbb C)$ is the space of $(k-1) \times (k-1)$ Hermitian matrices, is given by
\begin{equation} \label{eq:c-embedding}
    j([z]) = zz^*,~ z^*z =1.
\end{equation}
 This embedding is a $SU(k-1)$ equivariant embedding, where $SU(k-1)$ is the special unitary group $(k-1) \times (k-1)$ matrices of determinant 1, since $j([Az]) = Aj([z])A^*,, \forall A \in SU(k-1).$ The corresponding  extrinsic mean (set) of a random shape $X$ on $\mathbb{C}P^{k-2}$ is called the {\em VW mean (set)} (See Patrangenaru and Ellingson (2015), ch. 3 \cite{PaEl:2015}), and the {\em VW mean}, when it exists, and is labeled $\mu_{VW}(X), \mu_{VW}$ or simply $\mu_{E}.$ The corresponding  extrinsic antimean (set) of a random shape $X$, is called the {\em VW antimean (set)} and is labeled $\alpha\mu_{VW}(X), \alpha\mu_{VW}$ or $\alpha\mu_{E}.$

We have the following theorem for VW antimeans associated with the embedding \eqref{eq:c-embedding}.
\begin{theorem} \label{th:c-antimean}
Let $Q$ be a probability distribution on $\mathbb{C}P^{k-2}$ and let $\{[Z_{r}], \parallel Z_{r} \parallel = 1_{r = 1, \dots, n}\}$ be i.i.d.r.o.'s from $Q$. $(a)$ $Q$ is VW nonfocal iff $\lambda_1$, the smallest eigenvalue of $E[Z_{1}Z_{1}^{*}]$ is simple and in this case $\alpha \mu _{E}{Q} = [\nu],$ where $\nu$ is an eigenvector of $E[Z_{1}Z_{1}^{*}]$ corresponding to $\lambda_1$, with $\parallel \nu \parallel = 1$. $(b)$ The sample VW antimean $\alpha \overline{X}_{E} = {[m]}$, where $m$ is an eigenvector of norm 1 of $J = \frac{1}{n} \sum^n_{i=1}z_i z^*_i$, $\| z_i\| = 1, i = 1, \dots, n$, corresponding to the smallest eigenvalue of $J$, provided this eigenvalue has multiplicity one.
\end{theorem}

{\bf Proof.}
(a). The squared distance on the space $S(k-1,\mathbb{C})$  of Hermitian
matrices is $d_0^2(A,B) = Tr (( A-B)(A-B)^*) = Tr (( A-B)^2).$ A random object $X=[U]$ on $\mathbb C P^{k-2},$ with $U^*U = 1,$ has Fr\'echet function
\begin{equation}\label{eq:frechet-vw}
\mathcal F_X([u]) = E(Tr(UU^*-uu^*)^2), u^*u=1.
\end{equation}

The matrix $A = E(Tr(UU^*)$ is positive semidefinite, having the eigenvalues ${\lambda}_1 \geq {\lambda}_2 \geq \ldots \geq, {\lambda}_{k-1}\geq 0,$ and can be represented
as $A = B\Lambda B^*,$  where $B \in SU(k-1)$ and $\Lambda = \text{Diagonal}({\lambda}_1 \geq {\lambda}_2 \geq \ldots \geq, {\lambda}_{k-1}).$ From \eqref{eq:frechet-vw}, we get $ \mathcal F_X([u])= Tr(A^2) + 1 - 2 Tr (u^*Au),$ thus $F_X$ is maximized iff $ [u] \to Tr (u^*Au)$ is minimized, or $[v] \to Tr(v^*\Lambda v$ is minimized, where $v = Bu.$ Note that $v^*v=1,$ and if $v^T = (v_1 \dots v_{k-1}),$  then $Tr(v^*\Lambda v) = \sum_{a=1}^{k-2}\lambda_a|v_a|^2 \ge \lambda_{k-2} = \mathcal F_X([u_{k-1}]),$ where $u_{k-1}$ is an eigenvector of $A$ corresponding to the eigenvalue ${\lambda}_{k-1}.$ Part (b) follows, by taking the empirical distribution, with a matrix corresponding to the population expectation $A$ in Part (a) being given by
\begin{equation}
J = \hat{A} := n^{-1} \sum^n_{r=1} {\bf Z_r Z^*_r}.
\end{equation}

\subsection{Simulation} \label{sc:2.1}
\begin{figure}[ht!]
    \centering
    \begin{minipage}[b]{0.45\linewidth}
        \includegraphics[width=1.1\textwidth]{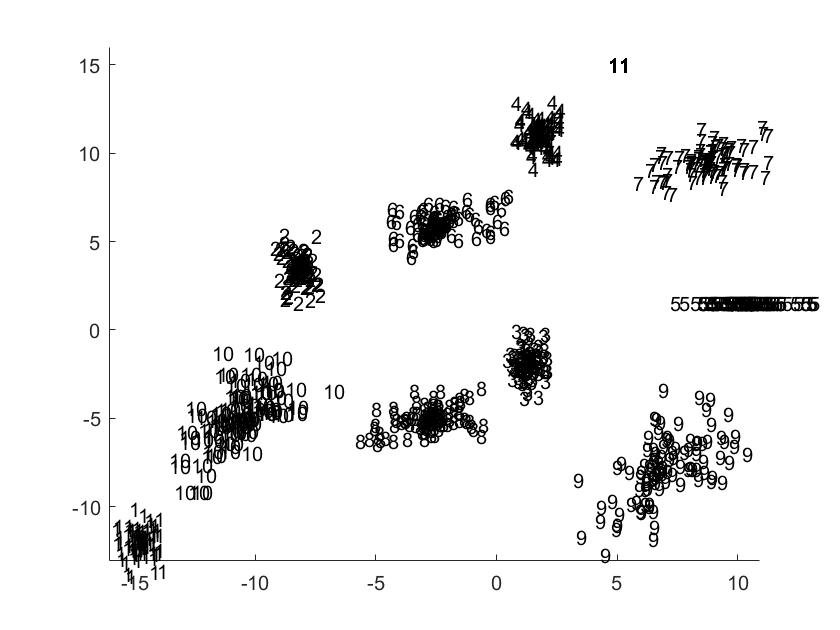}
        \caption{Simulated centered and scaled landmark configurations - affine coordinate representation}
        \label{f:s-scatter}
    \end{minipage}
    \quad
    \begin{minipage}[b]{0.45\linewidth}
        \includegraphics[width=1.1\textwidth]{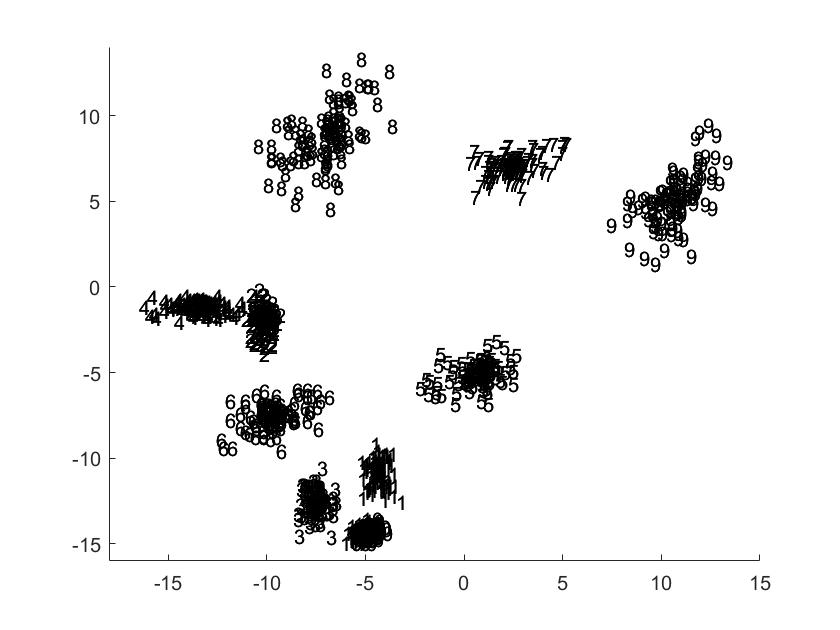}
        \caption{Simulated and Location removed landmark configurations}
        \label{f:simu}
    \end{minipage}
\end{figure}

We ran a simulation using an example of the VW embedding of a complex projective space ( a Kendall shape space) to compare VW means and VW antimeans for a data set of landmark configuration. In this context we ran a nonparametric bootstrap for sample VW means and sample VW antimeans. The objective of our simulations was to see if the bootstrap distributions of the sample VW means (respectively sample VW antimeans) is concentrated or not. For this simulations, the data represents coordinates of $k=11$ landmarks, and it has $N=100$ observations. The data are displayed in figure \ref{f:s-scatter}. Note that the corresponding shape variable is valued in $\mathbb C P^9$ (real dimension = 18).

Our study is on Kendall shape spaces, slightly more general than just using Bookstein coordinates (see Bookstein (1997)\cite{Bo:1997}) on this shape manifold. A useful tool  for ``removing location" of a $k$-ad, is the multiplication by a Helmert sub-matrix H, consisting in the last $(k - 1) \times k$ rows of a Helmert matrix. The full Helmert matrix HF, commonly used in Statistics, is a square $k \times k$ orthogonal matrix with its first row equal to $1/\sqrt{k}\mathbf{1}_k^T$, having the remaining rows orthogonal to the first row, with an increasing number of nonzero entries, as in \eqref{eq:helmert} . We drop the first row of HF so that the resulting matrix H does not depend on the original location of the configuration (see Dryden and Mardia(2016)\cite{DrMa:2016}). The $j$th row of the Helmert sub-matrix H is given by
            \begin{equation}\label{eq:helmert}
                (h_j, \cdots, h_j,-jh_j,0,\cdots,0),~ h_j = {\{j(j+1) \}}^{-1/2}.
            \end{equation}

\begin{figure}[ht!]
    \centering
    \begin{minipage}[b]{0.45\linewidth}
        \includegraphics[width=1.1\textwidth]{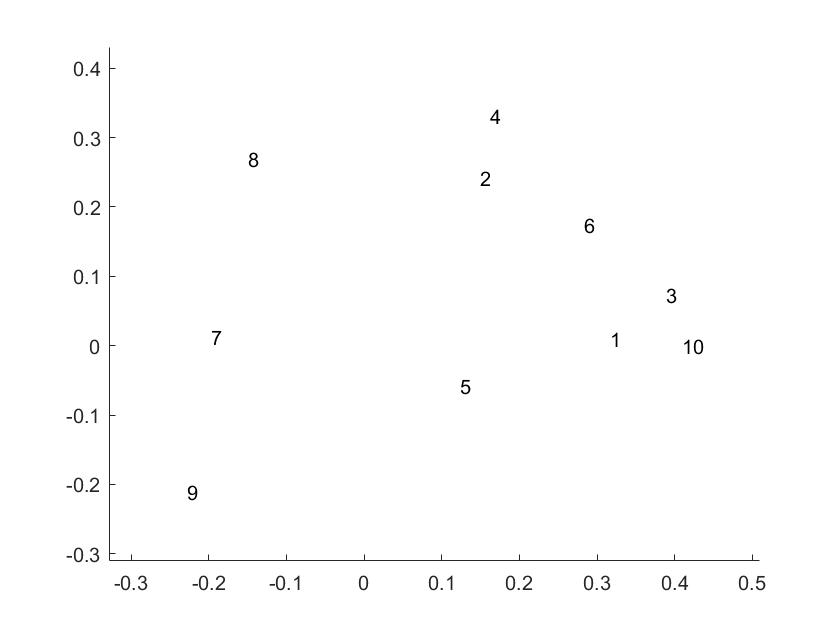}
        \caption{Icon of Sample VW mean shape of simulated landmark configurations}
        \label{f:s-mean}
    \end{minipage}
    \quad
    \begin{minipage}[b]{0.45\linewidth}
        \includegraphics[width=1.1\textwidth]{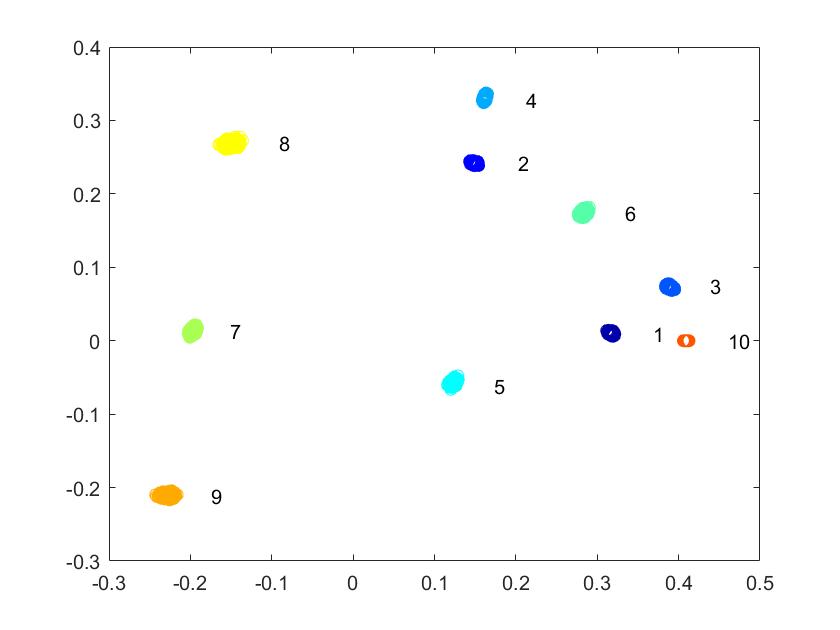}
        \caption{Distribution of sample VW means for bootstrap resamples}
        \label{f:s-bmean}
    \end{minipage}
\end{figure}

\begin{figure}[ht!]
    \centering
    \begin{minipage}[b]{0.45\linewidth}
        \includegraphics[width=1.1\textwidth]{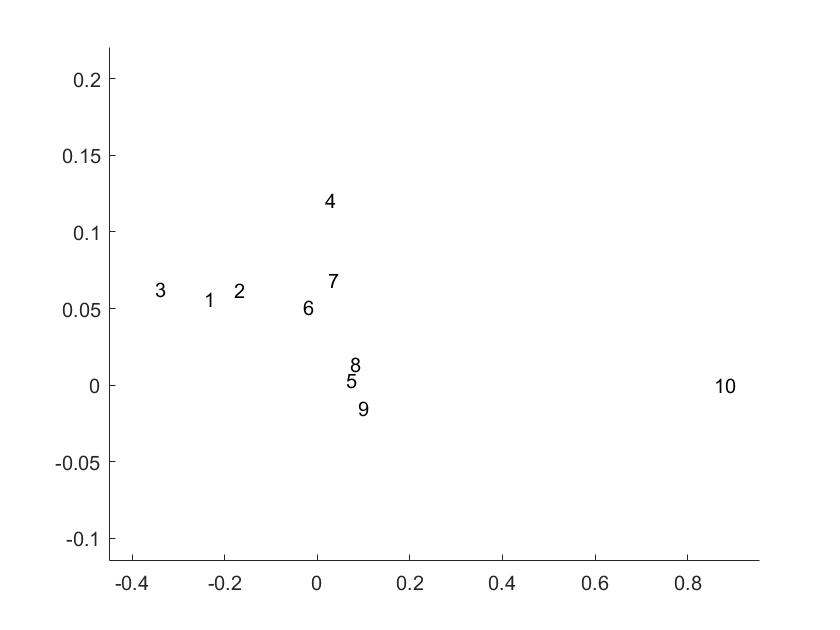}
        \caption{Icon of sample VW antimean shape of simulated landmark data}
        \label{f:s-antimean}
    \end{minipage}
    \quad
    \begin{minipage}[b]{0.45\linewidth}
        \includegraphics[width=1.1\textwidth]{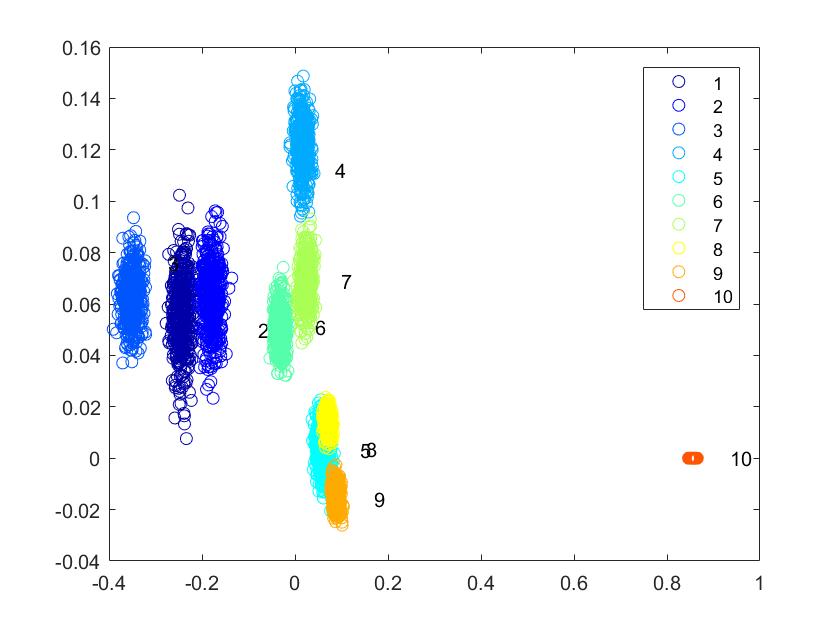}
        \caption{Distribution of icons of sample VW antimeans for bootstrap resamples.}
        \label{f:s-bantimean}
    \end{minipage}
\end{figure}

{To compute the sample VW mean or the sample VW antimean, we multiply by the Helmert sub-matrix H to ``remove location" of the original data. The Helmerized data after having removed location, is displayed in Figure \ref{f:simu}.}
The figure \ref{f:s-mean} is a representative (icon) of the sample VW mean of the coordinates of landmarks of the mean shape after removing location. {One may notice that the configurations in Figure \ref{f:simu} and Figure \ref{f:s-mean} look fairly similar, and the icon of the VW mean configuration is close, up to a rotation and scaling, to the icons of the sampled configurations. }
We computed the nonpivotal bootstrap distribution of the sample VW means in MATLAB, that we ran for 500 random resamples with repetition. An icon of the spherical representation of the bootstrap distribution of the sample VW means is displayed in Figure \ref{f:s-mean}. Note that the distribution of sample VW means bootstrap resamples is very concentrated around the sample VW means. As for the sample VW antimean shape, its representative is shown in Figure \ref{f:s-antimean}. The relative location of the landmarks in the icon of the sample VW antimean shape should look very different, when compared with the original landmark configuration, after the registration process, and indeed it does (see Figure \ref{f:s-antimean}).

We computed the nonpivotal bootstrap sample VW antimeans distribution using MATLAB, that we ran on 500 random resamples. Coordinates of the bootstrap distribution of the icons of the sample VW antimeans are displayed in Figure \ref{f:s-bantimean}. Note that the distribution of the landmark configuration for icons of the bootstrap sample VW antimeans are not too concentrated around the sample VW antimean; nonetheless they are similarly positioned.

From Theorem \ref{th:c-antimean}, in our simulation example, we know that the sample VW antimean is represented by an eigenvector of norm 1 of $J = \frac{1}{n} \sum^n_{i=1}z_i z^*_i$, $\| z_i\| = 1, i = 1, \dots, n$, corresponding to the smallest eigenvalue of J, where $[z_i] \in \mathbb CP^9$ are obtained by applying the submatrix of the last 10 rows of the Helmert matrix (see Mardia et al \cite{MaKeBi:1979}, p. 461) to the centered normalized data point $x_i+y_i \in \mathbb C^{11}.$ The smallest eigevalue of $J$ is very close to zero, since data is fairly concentrated, explaining the pattern in Figure  \ref{f:s-bantimean}.

\subsection{Application}
\begin{figure}[ht!]
    \centering
    \begin{minipage}[b]{0.45\linewidth}
        \includegraphics[width=1.1\textwidth]{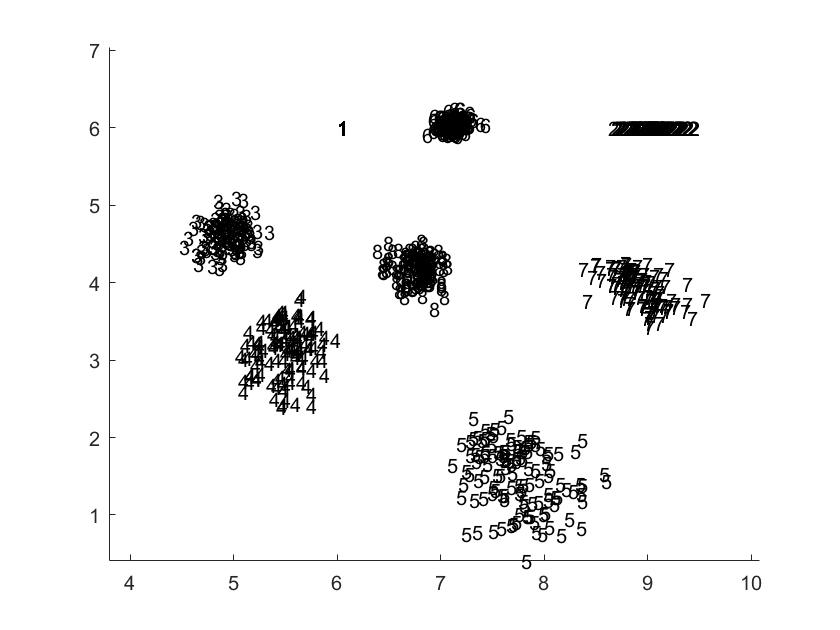}
        \caption{The coordinates of mid-sagital landmark configurations of midfaces of children skulls}
        \label{f:scatter}
    \end{minipage}
    \quad
    \begin{minipage}[b]{0.45\linewidth}
        \includegraphics[width=1.1\textwidth]{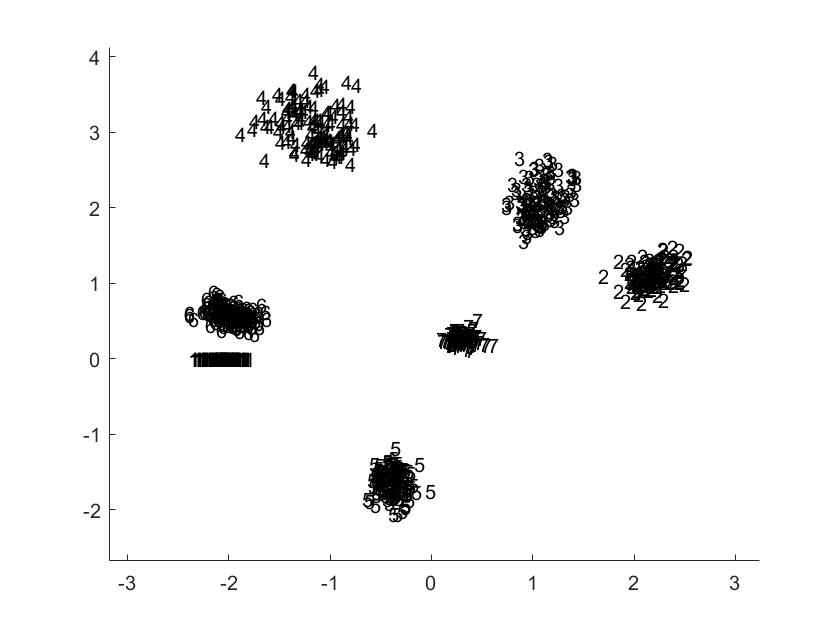}
        \caption{The coordinates of configurations in Figure 7 after location was removed}
        \label{f:child}
    \end{minipage}
\end{figure}

 We are interested to determine how concentrated is the bootstrap distribution  of the sample VW antimeans around the sample VW antimean, in the case of shapes of landmark configurations extracted from medical imaging outputs. Our data consists of shapes for a group of eighth midface anatomical landmarks labeled X-rays of skulls of eight year old and fourteen year-old North American children(72 boys and 52 girls), known as the {\em University School data}. The data set represents coordinates of landmarks, whose names and position on the skull are given in Bookstein ((1997)\cite{Bo:1997}). In Bhattacharya and Patrangenaru(2005) \cite{BP05} only part of this data set ( boys only) was used. The registered coordinates are displayed in Figure \ref{f:scatter}. The shape variable is valued in a Kendall space of planar octads, $\mathbb C P^6$ ( real dimension = 12 ).

\begin{figure}[ht!]
    \centering
    \begin{minipage}[b]{0.45\linewidth}
        \includegraphics[width=1.1\textwidth]{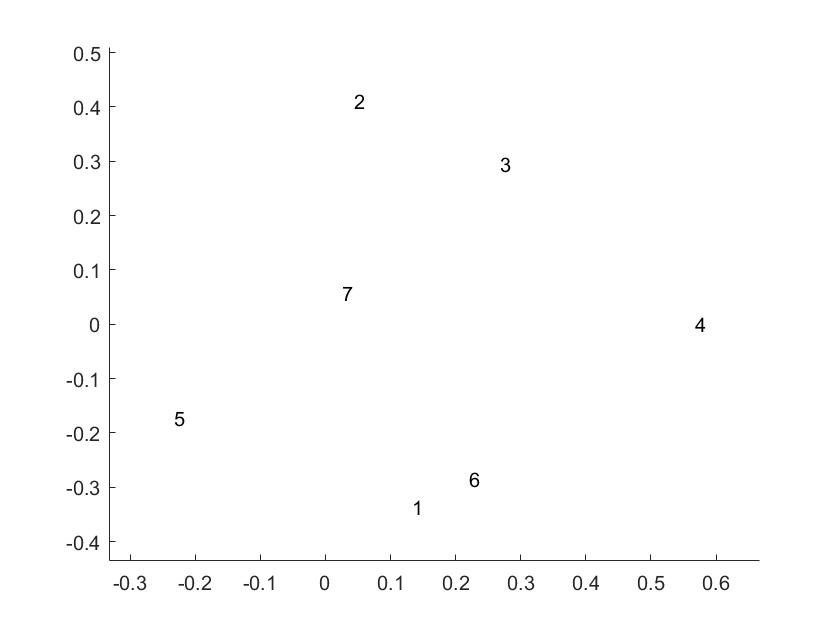}
        \caption{Icon of Helmertized sample VW mean shape of midface cranial landmark configurations}
        \label{f:mean}
    \end{minipage}
    \quad
    \begin{minipage}[b]{0.45\linewidth}
        \includegraphics[width=1.1\textwidth]{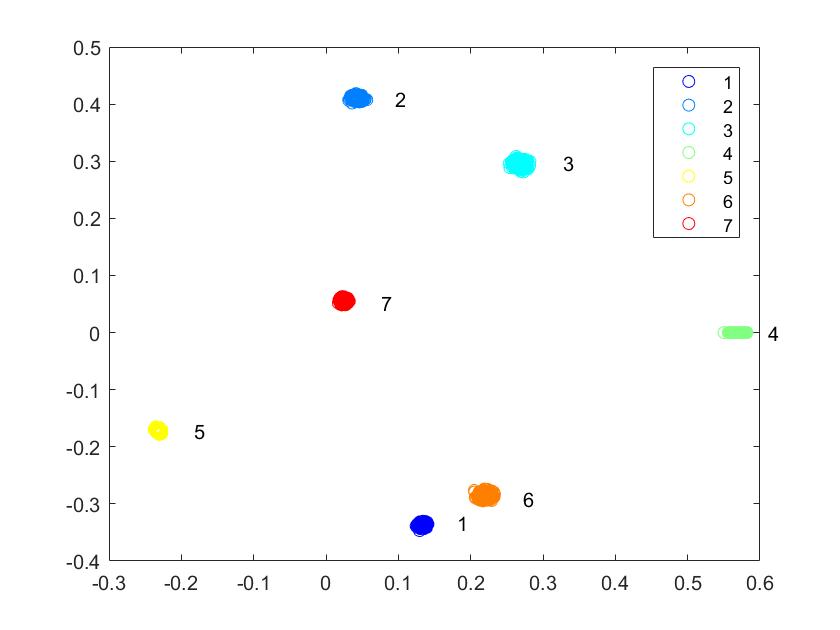}
        \caption{Distribution of icons of Helmertized bootstrap sample VW means}
        \label{f:bmean}
    \end{minipage}
\end{figure}

In our application, the data is registered using a Helmert sub-matrix $H$ in equation \eqref{eq:helmert}, with $k = 8,$ and is displayed in Figure \ref{f:child}.
In Figure \ref{f:mean} is displayed an icon of the sample VW mean of the Helmertized data, in a spherical representation. Like with the simulated data, one may notice that with VW mean icon, has a fairly close shape to the shapes of sampled configuration.

Next, we computed the nonparametric bootstrap distribution of the sample VW means shapes in MATLAB, that we ran for 500 random resamples. An icon of the Helmertized spherical representation of the bootstrap distribution of the sample VW means is displayed in Figure \ref{f:bmean}. Note that the bootstrap distribution of the sample VW means is very concentrated around the sample VW means, as theoretically predicted.

\begin{figure}[ht]
    \centering
    \begin{minipage}[b]{0.45\linewidth}
        \includegraphics[width=1.1\textwidth]{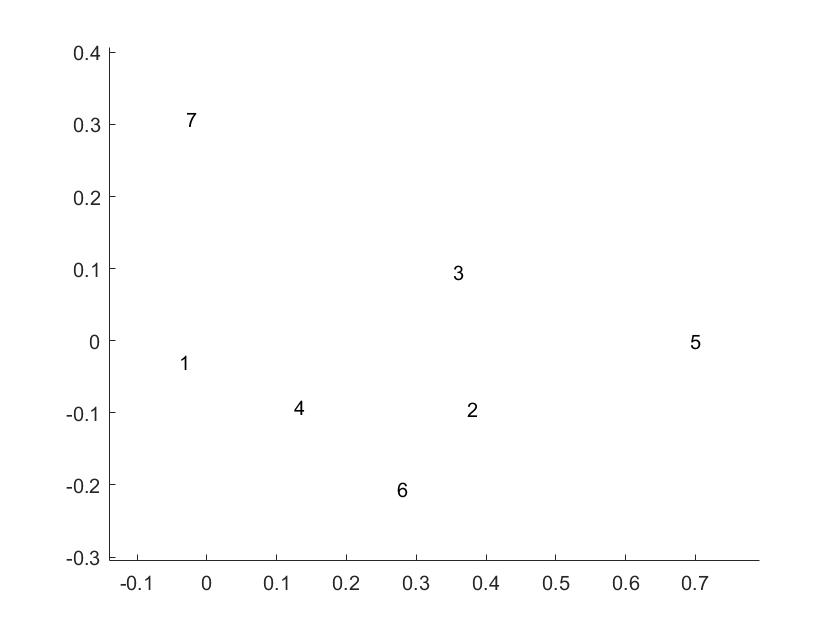}
        \caption {Icon of Helmertized sample VW antimean shape of midface cranial landmark configurations}
        \label{f:antimean}
    \end{minipage}
    \quad
    \begin{minipage}[b]{0.45\linewidth}
        \includegraphics[width=1.1\textwidth]{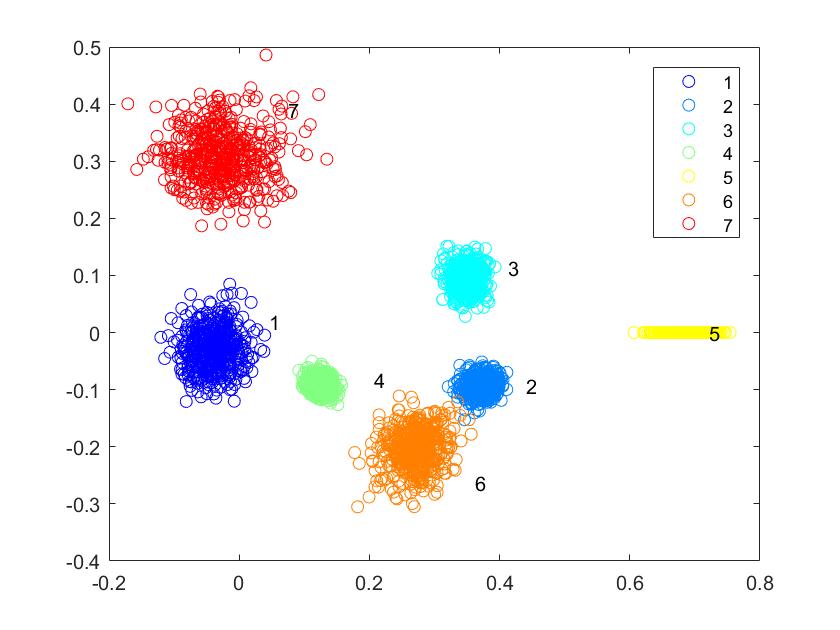}
        \caption{Distribution of icons of Helmertized bootstrap sample VW antimeans}
        \label{f:bantimean}
    \end{minipage}
\end{figure}

As for the sample VW antimean, its representative is shown in Figure \ref{f:antimean}. Since the sample VW antimean is on average far from the shape data, it is not surprising that the relative location of the landmarks in the sample VW antimean icon looks quite different from the one in the configurations in the original shape data.

We computed the nonparametric bootstrap distribution using MATLAB, that we ran again for 500 random resamples. A spherical representation of the bootstrap distribution of the sample VW antimeans in Helmetrized coordinates is displayed in Figure \ref{f:bantimean}. Here again, the icons of configurations for bootstrap distribution of the sample VW antimeans have a similar look with the one of the VW sample antimean, however is that concentrated around the registered icon of the sample VW antimean, partially due to computational rounding errors for eigenvectors associated with the smallest eigenvalue of $J^*.$ The standard affine embedding: $\mathbb{C}^{k-2} \to \mathbb{C}P^{k-2}$ is $(z^1, \cdots, z^{k-2}) \to [z^1: \cdots : z^{k-2} : 1]$, leads to the notion of affine coordinates of a projective point
\begin{equation}
    p = [z^1: \cdots : z^{k-1}], ~ z^{k-1} \neq 0
\end{equation}
{to be defined as}
\begin{equation}
    (w^1, w^2, \cdots, w^{k-2}) = (\frac{z^1}{z^{k-1}}, \cdots, \frac{z^{k-2}}{z^{k-1}}).
\end{equation}
Using simultaneous complex confidence intervals (See Bhattacharya and Patrangenaru 2005 \cite{BP05}) for the affine coordinates of the VW antimean, we obtain the following results: $w^1$ : [-0.1804 - 0.1808i   0.0549 + 0.1365i], $w^2$ : [0.4913 - 0.2301i   0.6136 - 0.0747i], $w^3$ : [0.4455 + 0.0385i   0.5885 + 0.2288i], $w^4$: [0.1344 - 0.1923i   0.2346 - 0.0748i], $w^5$: [0.2376 - 0.4823i   0.5682 - 0.1533i], $w^6$: [-0.2752 + 0.2558i   0.1936 + 0.8011i].

\section{VW anticovariance matrices and pivotal confidence regions for VW Antimeans }\label{s:3}
In this section we will discuss the asymptotic distribution of sample antimeans in axial data analysis and in planar shape analysis, after a review of a Central Limit Theorem for extrinsic sample antimeans.

\subsection{Central Limit Theorem for Extrinsic Sample Antimeans} \label{sc:3.1}
 In preparation, we are an using the large sample distribution for extrinsic sample antimeans given in Patrangenaru et al (2016 \cite{PG16}).

Assume $j$ is an embedding of a $d$-dimensional manifold $\mathcal M$ such that $j(\mathcal M)$ is closed in $\mathbb R^k$, and $Q = P_X$ is a $\alpha j$-nonfocal probability measure on $\mathcal M$ such that $j(Q)$ has finite moments of order 2. Let $\mu$ and $\Sigma$ be the mean and covariance matrix of $j(Q)$ regarded as a probability measure on $\mathbb R ^k$. Let $\mathcal F$ be the set of $\alpha j$-focal points of $j(\mathcal M)$, and let $P_{F,j}:{\mathcal F}^c \to j(\mathcal M)$ be the farthest projection on $j(\mathcal M)$. $P_{F,j}$ is differentiable at $\mu$ and has the differentiability class of $j(\mathcal M)$ around any $\alpha j$ nonfocal point.

A local frame field
$p \to (e_1(p), \dots, e_k(p))$, defined on an open neighborhood $U \subseteq \mathbb R^k$ is {\em adapted to the embedding $j$} if it is an orhonormal frame field and $\forall x \in j^{-1}(U), e_r(j(x)) = d_x j(f_r(x)), r= 1, \dots, d,$ where $(f_1, \dots, f_d)$ is a local frame field on $\mathcal M.$

Let $e_1, \dots, e_k$ be the canonical basis of $\mathbb R^k$ and assume $(e_1(p), \dots, e_k(p))$ is an adapted frame field around $P_{F,j} (\mu) =  j(\alpha \mu_E)$. Then $d_\mu P_{F,j} (e_b) \in T_{P_{F,j}(\mu)}j(\mathcal M)$ is a linear combination of $e_1(P_{F,j} (\mu)), \dots, e_d(P_{F,j} (\mu))$:
\begin{equation}
    d_\mu P_{F,j} (e_b) = \sum^d_{a=1} (d_\mu P_{F,j} (e_b)) \cdot e_a(P_{F,j} (\mu))e_a(P_{F,j} (\mu))
\end{equation}
By the delta method, $n^{1/2}(P_{F,j} (\overline{j(X)}) - P_{F,j}(\mu))$ converges weakly to $N_k(0_k,\alpha \Sigma_\mu)$, where $\overline{j(X)} = \frac{1}{n} \sum^n_{i=1} j(X_i)$ and
\begin{equation}
    \begin{aligned}
       \alpha \Sigma_\mu = [\sum^d_{a=1} d_\mu P_{F,j} (e_b) \cdot e_a(P_{F,j} (\mu))e_a(P_{F,j} (\mu))]_{b=1,\dots,k} \\ \times \Sigma [\sum^d_{a=1} d_\mu P_{F,j} (e_b) \cdot e_a(P_{F,j} (\mu))e_a(P_{F,j} (\mu))]^T_{b=1,\dots,k}
    \end{aligned}
\end{equation}
here $\Sigma$ is the covariance matrix of $j(X_1)$ w.r.t the canonical basis $e_1, \dots, e_k$.

The asymptotic distribution $N_k(0_k,\alpha \Sigma_\mu)$ is degenerate and the support of this distribution is on $T_{P_{F,j}}j(\mathcal M)$, since the range of $d_\mu P_{F,j}$ is $T_{P_{F,j(\mu)}}j(\mathcal M)$. Note that $d_\mu P_{F,j} (e_b) \cdot e_a(P_{F,j} (\mu)) = 0$ for $a=d+1, \dots,k$.

The tangential component $tan(v)$ of $v \in \mathbb R^k$, w.r.t the basis $e_a(P_{F,j}(\mu)) \in T_{P_{F,j(\mu)}}j(\mathcal M), a = 1, \dots, d$ is given by
\begin{equation}
    tan(v) = [e_1{(P_{F,j} (\mu))}^T v, \dots, e_d{(P_{F,j} (\mu))}^T v]^T
\end{equation}
Then the random vector ${(d_{\alpha \mu_{E}}j)}^{-1}(tan(P_{F,j} (\overline{(j(X))})-P_{F,j}(\mu))) = \sum^d_{a=1} {\overline{X}}^a_j f_a$ has the following covariance matrix w.r.t the basis $f_1(\alpha \mu _E), \dots, f_d(\alpha \mu_E)$:
\begin{equation}
    \begin{aligned}
        \alpha \Sigma_{j,E} =   e_a{(P_{F,j} (\mu))}^T \alpha \Sigma_\mu e_b{(P_{F,j} (\mu))}_{1 \leq a, b \leq d} \\ = [\Sigma d_\mu P_{F,j} (e_b) \cdot e_a(P_{F,j} (\mu))]_{a=1,\dots,d} ~ \Sigma  ~  [\Sigma d_\mu P_{F,j} (e_b) \cdot e_a(P_{F,j} (\mu))]^T _{a=1,\dots,d}
    \end{aligned}
\end{equation}
which is the {\em anticovariance} matrix of the random object $X$. Similarly, given i.i.d.r.o.'s $X_1, \cdots, X_n$ from $Q$, we define the {\em sample anticovariance} matrix $aS_{j,E,n}$ as the anticovariance matrix associated with the empirical distribution $\hat{Q}_n.$

\subsection{VW anticovariance in $\mathbb R P ^{N-1}$ and $\mathbb C P^{k-2}$} \label{sc:3.2}
We first consider the case when $\mathcal M = \mathbb R P^{N-1}$, the real projective space which can be identified with the sphere $S^{N-1} = \{x \in \mathbb R ^N | {\|x \|}^2 =1 \}$ with antipodal points identified(see Mardia and Jupp (2009)\cite{MaJu:2009} ). Here the points in $\mathbb R^{N}$ are regarded as $N \times 1$ vectors. $\mathbb R P^{N-1}$ can be identified with the quotient space $S^{N-1}/\{x, {-x} \}$; it is a compact homogeneous space, with the group $SO(N)$ acting transitively on ($\mathbb R P^{N-1}, \rho_0$), where the distance $\rho_0$ on $\mathbb R P^{N-1}$ is induced by the chord distance on the space $S(N, \mathbb R)$ of symmetric $N \times N$ and the embedding $j$ that is compatible with two transitive group actions of $SO(N)$ on $\mathbb R P^{N-1}$, respectively on $j(\mathbb R P^{N-1})$, that is
\begin{equation}
j(T \cdot [x])= T \otimes j([x]), ~~\forall~T\in SO(N),~~ \forall~[x] \in \mathbb{R}P^{N-1}
\end{equation}
where $T \cdot [x] = [Tx]$ and $T \otimes A$ is given in \eqref{otimes} below.

Such an embedding is said to be {\em equivariant} (See Kent 1992 \cite{Ke:1992}). The {\em equivariant} embedding of $\mathbb{R}P^{N-1}$ that was used so far in the axial data analysis literature is the Veronese Whitney (VW) embedding $j:\mathbb{R}P^{N-1} \to S_+(N, \mathbb{R})$, that associates to an axis the matrix of the orthogonal projection on this axis (See Patrangenaru and Ellingson 2015 \cite{PaEl:2015}, chapter 3)
\begin{equation} \label{eq:R-embedding}
j([x])= xx^{T}, \|x\|=1
\end{equation}
Here $S_+(N, \mathbb R)$ is the set of nonnegative definite symmetric $N \times N$ matrices, and in this case
\begin{equation}\label{otimes}
    T \otimes A = TAT^T, ~~\forall~T\in SO(N),~~\forall~A\in S_+(N, \mathbb R)
\end{equation}

\begin{definition}
A random object $[X]=Y$ on $\mathbb R P^{N-1}$ is $\alpha$VW-nonfocal if it is $\alpha j$-nonfocal w.r.t. the VW embedding in (\ref{eq:R-embedding}).
\end{definition}

Then we have the following proposition from Patrangenaru et al (2016) \cite{PaYagu:2016}.
\begin{proposition} \label{pro:nonfocal}
A random object $[X]=Y$ on $\mathbb R P^{N-1}$, $X^TX=1$ is $\alpha$VW-nonfocal iff the smallest eigenvalue of $E(XX^T)$ is positive and has multiplicity 1.
\end{proposition}

Now we consider the anticovariance on $\mathbb R P^{N-1}$.
\begin{proposition}\label{pro:anticov-real}
Assume [$X_r$], $X_r^T X_r = 1, r = 1, \dots, n$ is a random sample from a $\alpha j$-nonfocal probability measure $Q$ on $\mathbb{R}P^{N-1}$. And $\lambda_a, a = 1, \dots, N$, are eigenvalues of $K:= \frac{1}{n} \sum ^n _{r=1} X_r X^T_r$ in increasing order and $m_a, a = 1, \dots, N$, are corresponding linearly independent unit eigenvectors. Then the sample VW anticovariance matrix $a S_{j,E,n}$ is given by
\begin{equation}
    (a S_{j,E,n})_{ab} = n^{-1}{(\lambda _a - \lambda _1)}^{-1}{(\lambda_b-\lambda_1)}^{-1} \times \sum_r(m_a \cdot X_r)(m_b \cdot X_r){(m_1 \cdot X_r)}^2
\end{equation}
\end{proposition}

The {\bf proof} is along the lines of a similar result from sample VW covariance on $\mathbb R P^{N-1}$ (see Bhattacharya and Patrangenaru (2003)\cite{BhPa:2003}).

As the embedding $j$ is equivariant, w.l.o.g. we may assume that the farthest projection of sample mean $j(a {\bar{X}}_{j,E}) = P_{F,j}(\overline{j(X)})$ is a diagonal matrix, $a {\bar{X}}_{j,E} = [m_1] = [e_1]$ and the other unit eigenvectors of $\overline{j(X)} = D$ are $m_a = e_a$, $\forall a = 2, \dots, N$. Based on this description of tangent space at $[x]$ we evaluate $d_D P_{F,j}$. $T_{[x]} \mathbb R P^{N-1}$, one can select  the orthonormal frame $e_a(P_{F,j}(D)) = d_{[e_1]}j(e_a)$ in $T_{P_{F,j}}(D)j(\mathbb R P ^{N-1})$. Note that $S(N,R)$ has the orthobasis $F^b_a, b \leq a$, where, for $a < b$, the matrix $F^b_a$ has the positions $(a,b)$, $(b,a)$ that are equal to $2^{-1/2}$, and all other entries equal to zero. We note that also $F^b_a=j([e_a])$.

$d_D P_{F,j}(F^b_a) = 0$, $\forall a,b, 2 \leq b <a \leq N$ and $d_D P_{F,j}(F^b_1) = (\lambda_b - \lambda_1)^{-1} e_a(P_{F,j}(D))$; If $\lambda_a, a = 1, \dots, N$ are the eigenvalues of $D$ in their increasing order. Then from this equation it follows that, if $\overline{j(X)}$ is a diagonal matrix $D$, then the entry $(a S_{j,E,n})_{ab}$ is given by.
$$(a S_{j,E,n})_{ab} = n^{-1}(\lambda_b - \lambda_1)^{-1}(\lambda_a - \lambda_1)^{-1} \sum _{r=1}^n X^a_r X^b_r {(X^r_1)}^2$$
Taking $\overline{j(X)}$ to be a diagonal matrix and $m_a = e_a$

Let $T([\nu]) = n {\|a S_{E,n}^{- \frac{1}{2}} tan (P_{F,j}(\overline{j(X)}) - P_{F,j}(E(j(Q)))) \|}^2$ (see formula (27) from Patrangenaru et al 2016 \cite{PG16}). We can derive now the following theorem. Note that $\alpha \mu_{j,E} = [\nu_1]$, where $(\nu_a), a = 1, \dots, N$, are unit eigenvectors of $E(XX^t) = E(j(Q))$ corresponding to eigenvalues in their increasing order.

\begin{theorem}
Assume $j$ is the Veronese-Whitney embedding of $\mathbb{R} P^{N-1}$ and let [$X_r$], $X_r^T X_r = 1, r = 1, \dots, n$ be a random sample from an $\alpha j$-nonfocal distribution $Q$. Then $T([\nu])$ is given by
$$T([\nu]) = n \nu ^t[{(\nu _a)}_{a = 2, \dots, N}]a S_{E,n}^{-1}{[{(\nu _a)}_{a = 2, \dots, N}]}^t \nu,$$
and, asymptotically, $T([\nu])$ has a $\chi _{N-1} ^2$ distribution.
\end{theorem}

Proof: Let $\nu_2, \dots, \nu_{N}$ be the orthobasis of the tangent space $T_{[\nu_1]} \mathbb R P ^{N-1}$. Based on that the VW embedding $j$ is isometric and using the method of moving frame(See Bhattacharya and Patrangenaru 2005 \cite{BP05}). Let $e_a(P_{F,j}(\nu_{E,j})) = (d_{[\nu_1]j})(\nu_a)$ be the first elements of the adapted moving frame. Then the $a$th tangential component of $P_{F,j}(\overline{(j(X))})-P_{F,j}(\nu)$ w.r.t. this basis of $T_{P_{F,j}(E(j(Q)))}j(\mathbb{R}P^{N-1})$ equals up to a sign the $a$th component of $m-\nu_1$ w.r.t. the orthobasis $\nu_2, \dots, \nu_{N}$ in $T_{[\nu_1]} \mathbb R P ^{N-1}$, namely $\nu_a^t m$. 

We say that a random object on  $\mathbb C P^{k-2}$ is $\alpha$VW-nonfocal if it is $\alpha j$-nonfocal w.r.t. the embedding in (\ref{eq:c-embedding})

Similarly with Proposition \ref{pro:nonfocal}, we get the following proposition.
\begin{proposition}
The random object $Y=[Z]$ on $\mathbb C P^{k-1}$ is $\alpha$VW-nonfocal iff the smallest eigenvalue of $E \frac{ZZ^*}{Z^*Z}$ is positive and has multiplicity 1.
\end{proposition}

Similar asymptotic results can be obtained for the large sample distribution of VW means of planar shapes, as following. Recall that the planar shapes space $M = \Sigma ^k_2$ of an ordered set of $k$ points in $\mathbb C$ at least two of which are distinct, can be identified in different ways with the complex projective space $\mathbb C P ^{k-2}$ (see Bhattacharya and Patrangenaru (2003) \cite{BhPa:2003} ,and Balan and Patrangenaru(2005) \cite{BaPa:2005}). Here we regard $\mathbb C P ^{k-2}$ as a set of equivalence classes $\mathbb C P ^{k-2} = S^{2k-3} / S^1$ where $S^{2k-3}$ is the space of complex vectors in $\mathbb C ^{k-1}$ of norm 1, and the equivalence relation on $S^{2k-3}$ is by multiplication with scalars in $S^1$. The action of $S^1$ on $S^{2k-3}$ is by multiplication of complex vectors with scalars in $S^1$ (complex numbers of modulus 1). A complex vector $z = (z_1, z_2, \dots, z_{k-1})$ of norm 1 corresponding to a given configuration of $k$ landmarks, with the identification described in Bhattacharya and Patrangenaru ((2003) \cite{BhPa:2003}), can be displayed in the Euclidean plane (complex line) with the superscripts as labels.

A random variable $X=[Z],$ $\|Z\|=1,$ valued in $\mathbb C P^{k-2}$ is $\alpha j$-nonfocal if the smallest eigenvalue of $E[ZZ^*]$ is simple, and then the VW-antimean of $X$ is $\alpha \mu_{j,E} = [\nu]$, where $\nu \in \mathbb C^{k-1},$ $\| \nu \|=1$, is an eigenvector corresponding to this eigenvalue. The sample VW-antimean $\alpha \overline{[z]}_{j,E}$ of a random sample $[z_r]=[(z_r^1, \cdots, z_r^{k-1})]$, $\| z_r \|=1, r= 1, \cdots, n,$ from such a nonfocal distribution exists with probability converging to 1 as $n \to \infty$, and is the same as that given by
\begin{equation}
    a \overline{[z]}_{j,E} =[m],
\end{equation}
where $m$ is the unit eigenvector of
\begin{equation}\label{eq:K}
    K:=\frac{1}{n} \sum_{r=1}^n z_r z_r^*.
\end{equation}
corresponding to the smallest eigenvalue.

\begin{proposition}
Assume $X_r = [Z_r]$, $Z_r^T Z_r = 1, r = 1, \dots, n$ is a random sample from a $\alpha j$-nonfocal probability measure $Q$ with a nondegenerate $\alpha j$-extrinsic anticovariance matrix on $\mathbb C P^{k-2}$. $\lambda_a, a =2, \cdots, k-1$ are eigenvalues of K in \eqref{eq:K} in their increasing order and $m_a, a = 2, \cdots, k-1$ are corresponding linearly independent unit eigenvectors. Then the VW-extrinsic sample anticovariance matrix $ a S_{E,n}$ as a complex matrix has the entries
\begin{equation} \label{eq:anti-covariance}
    {(a S_{E,n})}_{ab}= n^{-1}{(\lambda_a - \lambda_1)}^{-1}{(\lambda_b - \lambda_1)}^{-1} \times \sum_{r=1}^n (m_a \cdot Z_r){(m_b \cdot Z_r)}^*{|m_1 \cdot Z_r|}^2
\end{equation}
\end{proposition}

The {\bf proof} is based on the equivariance of the VW embedding. The embedding $j$: $\mathbb C P^{k-2} \to S(k-1, \mathbb C)$, where the action $SU(k-1)$ is a non-negative semi defined self-adjoint complex matrices(see Bhattacharya and Patrangenaru(2003) \cite{BhPa:2003}). First we need to assume that $K:=\frac{1}{n} \sum_{r=1}^n z_r z_r^*$ is a diagonal matrix, the smallest eigenvalue corresponding complex eigenvector of norm 1 of $K$ is a simple root of the characteristic polynomial over $\mathbb C$, with $m_1=e_1.$

The tangent space $T_{[m_{1}]}j(\mathbb C P^{k-2})$ has an orthobasis $m_a^{'} =i e_a, ~ a= 2, \cdots, k-1$, where $m_a=e_a$ are eigenvector corresponding to the largest eigenvalue. Here we define a path $\eta_z(t)=[cos~tm_1 + sin~tz]$, where $z$ is orthogonal to $m_1 \in \mathbb C^{k-1}.$ $T_{P_{F,j}(K)}j(\mathbb C P^{k-2})$ is generated by the vectors tangent to such paths $\eta_z(t)$ at $t=0.$ Such a vector, has the form $zm_1^* + m_1z^*$, as a matrix in $S(k-1, \mathbb C)$.

Thus we take $z=m_a,~ a= 2, \cdots, k-1$, or $z=im_a,~ a= 2, \cdots, k-1$ based on the eigenvectors of $K$ are orthogonal w.r.t. the complex scalar product. We norm these vectors to have unit length to obtain the orthonormal frame.

       $$ e_a(P_{F,j}(K)) = d_{[m_1]}j(m_a) = 2^{-1/2}(m_am_1^* + m_1m_a^*),$$
       $$ e_a^{'} (P_{F,j}(K)) = d_{[m_1]}j(m_a) = i 2^{-1/2}(m_am_1^* + m_1m_a^*).$$

As we assume $K$ is diagonal. In this case $m_a = e_a,~ e_a(P_{F,j}(K)) = 2^{-1/2}E_{1}^{a}$ and $e_a^{'}(P_{F,j}(K)) = 2^{-1/2}F_{1}^{a}$, where $E_a^b$ has the positions $(a,b)$ and $(b,a)$ that are equal to 1 and all other entries zero, and $F_a^b$ has all the positions $(a,b)$ and $(b,a)$ that are equal to $i$, respectively $-i$ and other entries zero. That we have $d_K P_{F,j}(E_a^b)=d_K P_{F,j}(F_a^b)=0, ~ \forall 1 < a \leq b \leq k-1$, and
        $$d_K P_{F,j}(E_1^a)={(\lambda_a - \lambda_1)}^{-1}e_a(P_{F,j}(K)), $$
        $$d_K P_{F,j}(F_1^a)={(\lambda_a - \lambda_1)}^{-1}e_a^{'}(P_{F,j}(K)).$$
We evaluate the extrinsic sample anticovariance matrix $aS_{E,n}$ in formula (25) in Patrangenaru et al (2016 \cite{PG16}) using the real scalar product in $S(k-1, \mathbb C)$, namely, $U \cdot V = ReTr(UV^*)$. Note that,
        $$d_K P_{F,j}(E_1^b) \cdot e_a(P_{F,j}(K)) = {(\lambda_a - \lambda_1)}^{-1} \delta _{ba}, $$
        $$d_K P_{F,j}(E_1^b) \cdot e_a ^{'} (P_{F,j}(K)) = 0$$
and
       $$ d_K P_{F,j}(F_1^b) \cdot e_a^{'} {(P_{F,j}(K))}^T = {(\lambda_a - \lambda_1)}^{-1} \delta _{ba},$$
       $$ d_K P_{F,j}(F_1^b) \cdot e_a  (P_{F,j}(K)) = 0$$

Thus we may regard $aS_{E,n}$ as a complex matrix noting that in this case we get
\begin{equation}
    {(a S_{E,n})}_{ab}= n^{-1}{(\lambda_a - \lambda_1)}^{-1}{(\lambda_b - \lambda_1)}^{-1}  \sum_{r=1}^n (e_a \cdot Z_r){(e_b \cdot Z_r)}^*{|m_1 \cdot Z_r|}^2
\end{equation}
Thus proving (\ref{eq:anti-covariance}) when $K$ is diagonal. The general case follows by equivariance.

Next we consider the statistic
$$ T(a\bar{X}_E, \alpha \mu_E) = n {\|{(aS_{E,n})}^{-1/2} tan(P_{F,j}(\bar{j(X)})-P_{F,j}(\mu_{j(X_1)})\|}^2$$
given in Patrangenaru et al 2016 \cite{PG16}, in the our context of i.i.d.r.o objects on a complex projective space to get:

\begin{theorem} \label{th:quantity}
Let $X_r = [Z_r]$, $Z_r^T Z_r=1, ~ r=1, \cdots, n,$ be a random sample from a VW-$\alpha$nonfocal probability measure $Q$ on $\mathbb C P ^{k-2}$. Then the random variable given by
\begin{equation}\label{eq:t2-anti}
    T([m],[\nu]) = n[{(m \cdot \nu_a)}_{a = 2, \cdots, k-1}]{(aS_{E,n})}^{-1}{[{(m \cdot \nu_a)}_{a = 2, \cdots, k-1}]}^*
\end{equation}
has asymptotically a $\chi_{2k-4}^2$ distribution.
\end{theorem}
{\bf Proof}. Since the VW embedding $j$ is by definition isometric, and $(\nu_2, \cdots, \nu_{k-1}, \nu_2^*, \cdots \nu_{k-1}^*)$ is an orthogonal basis in the tangent space $T_{[\nu_1]} \mathbb C P^{k-2},$ the first elements of the adapted orthogonal moving frame are $e_a(P_j(\mu))=(d_{[\nu_1]}j)(\nu_a)$  $e_a^*(P_j(\mu))=(d_{[\nu_1]}j)(\nu_a^*).$ Then the $a$th tangential component of $P_{F,j}({j([m])})-P_{F,j}(\mu_{j(X_1)})$ w.r.t. this basis of $T_{P_j(\mu)} \mathbb C P^{k-2}$ equals up to a sign to the component of $m-\nu_1$ w.r.t. the orthobasis $\nu_2, \cdots, \nu_{k-1}$ in $T_{[\nu_1]} \mathbb C P^{k-2}$, which is $\nu_a^t m$; and the $a^*$th tangential components are given by ${\nu_a^*}^t m$, and together(in complex multiplication) they yield the complex vector $[{(m \cdot \nu_a)}_{a = 2, \cdots, k-1}]$. The result follows by taking $[m]=P_{F,j}(\bar{j(X)})=j(a\bar{X}_E).$

We may derive the following large sample confidence regions for the VW-antimean shape

\begin{corollary}
    Assume $X_r = [Z_r]$, $Z_r^T Z_r = 1, ~ r=1, \cdots, n,$ is a random sample from a $\alpha j$-nonfocal probability measure $Q$ on $\mathbb CP^{k-2}$. An asymptotic $(1-\beta)-$confidence region for $\alpha \mu_E^j (Q) = [\nu]$ is given by $R_\beta (X)=\{[\nu]: T([m],[\nu]) \leq \chi_{2k-4,\beta}^2 \}$, where $T([m],[\nu])$ is given in \eqref{eq:t2-anti}. If $Q$ has a nonzero absolutely continuous component w.r.t. the volume measure on $\mathbb CP^{k-2}$, then the coverage error of $R_\alpha (X)$ is of order $O(n^{-1})$.
\end{corollary}

When the sample size is small, the coverage error could be quite large, and a bootstrap analogue of Theorem \ref{th:quantity} is preferred.

\begin{theorem}
Let $X_r = [Z_r]$, $Z_r^T Z_r =1, ~ r=1, \cdots, n,$ be a random sample from a $\alpha VW$-nonfocal distribution $Q$ on $\mathbb C P ^{k-2}$, such that $X_r$ has a nonzero absolutely continuous component w.r.t. the volume measure on $\mathbb C P ^{k-2}$. If $j$ is the VW embedding, and the restriction of the covariance matrix of $j(X_1)$ to $T_{[\nu]} j(\mathbb C P ^{k-2})$ is nondegenerate, where $\alpha \mu_E (Q) = [\nu]$ be the extrinsic antimean of $Q$. For a bootstrap resample ${\{X_r^*\}}_{r=1. \cdots, n}$ from the given sample, consider the matrix $K^*:=n^{-1}\sum Z_r^* {Z_r^*}^*$. Let ${(m_a^*)}_{a=1, \cdots, k-1}$ be the unit complex eigenvectors, corresponding to the eigenvalues ${(m_a^*)}_{a=1, \cdots, k-1}$ in increasing order. Let ${(a S_{E,n})}^*$ be the matrix obtained from $a S_{E,n}$ by substituting all the entries with $*$-entires. Then the bootstrap distribution function of
\begin{equation}
    T({[m]}^*,[m]) = n[{(m_1^* \cdot m_a)}_{a = 2, \cdots, k-1}]{(aS^*_{E,n})}^{-1}{[{(m_1^* \cdot m_a)}_{a = 2, \cdots, k-1}]}^*
\end{equation}
approximates the true distribution function of $T([m],[\nu])$ given in Theorem \ref{th:quantity} with an error of order $O_P(n^{-2})$.
\end{theorem}


\begin{thebibliography}{99}

\bibitem{BaPa:2005} V. Balan, and V. Patrangenaru (2005). {\em Geometry of shape spaces},Proc. of The 5-th Conference of Balkan Society of Geometers.

\bibitem{BeFi:1998} R. Beran and N. I. Fisher (1998). {\em Nonparametric comparison of mean directions or mean axes}, Annals of statistics, 472--493.

\bibitem{BhPa:2003} R. N. Bhattacharya and V. Patrangenaru (2003). Large
sample theory of intrinsic and extrinsic sample means on manifolds-Part I,{\it
Ann. Statist.} {\bf 31}, no. 1, 1--29.

\bibitem{BP05}R.~N. Bhattacharya and V. Patrangenaru(2005). {\em Large sample theory  of intrinsic and extrinsic
    sample means  on manifolds}, Part II, Ann. Statist. \textbf{33}, 1211--1245.

\bibitem{Bo:1997} F. L. Bookstein (1997). {\em Morphometric tools for landmark data: geometry and biology}, Cambridge University Press.

\bibitem{DrMa:2016}  Dryden, Ian L. and Mardia, Kanti V. (2016). {\em Statistical shape analysis with applications in R. Second edition.}  Wiley Series in Probability and Statistics. John Wiley $\&$ Sons, Ltd., Chichester.

\bibitem{FiHaJiWo:1996} N. Fisher, P. Hall, B.-Y. Jing and  A.T.A. Wood (1996). {\em Improved pivotal methods for constructing confidence regions with directional data}, Journal of the American Statistical Association, {\bf 91}, 435, 1062--1070.

\bibitem{Fr:48}M. Fr\'echet(1948). {\em Les \'elements al\'eatoires de nature quelconque dans un espace distanci\'e}, Ann. Inst. H. Poincar\'e {\bf 10}, 215--310.

\bibitem{HeLa:1998}  H. Hendriks and Z. Landsman (1998). Mean location and sample mean location on manifolds: asymptotics, tests, confidence regions. {\em J. Multivariate Anal.} {\bf 67}, 227-â€“243

\bibitem{Ke:1984}D. G. Kendall (1984). {\em Shape manifolds, procrustean metrics, and complex projective spaces}. Bulletin of the London Mathematical Society, Oxford University Press.

\bibitem{Ke:1992} J.T. Kent (1992). {\em New directions in shape analysis. The Art of Statistical Science, A Tribute to GS Watson, 115--127},Wiley Ser. Probab. Math. Statist. Probab. Math. Statist., Wiley, Chichester.

\bibitem{MaJu:2009} K. V. Mardia and P. E. Jupp (2009) {\em Directional statistics}, John Wiley \& Sons

\bibitem{MaKeBi:1979} K.V. Mardia, J.T. Kent and J.M. Bibby (1979). {\em Multivariate Analysis}. Academic Press.

\bibitem{MaPa:2005} K. V. Mardia and V. Patrangenaru (2005). Directions and
projective shapes. {\em The Annals of Statistics}, {\bf 33}, 1666--1699.

\bibitem{PaEl:2015}V. Patrangenaru and L. Ellingson(2015), {\em Nonparametric Statistics on Manifolds and Their
    Applications to Object Data Analysis}, CRC-Chapman \& Hall.

\bibitem{PaLiSu:2010}V. Patrangenaru, X. Liu and S. Sugathadasa(2010). {\em Nonparametric 3D projective shape estimation
    from pairs of 2D images - I, In memory of W.P. Dayawansa}, Journal of Multivariate Analysis {\bf 101}, 11--31.

\bibitem{PaMa:2003} V. Patrangenaru and K. V. Mardia (2003). Affine shape analysis and image analysis, {\em Proceedings of the Leeds Annual
Statistics Research Workshop.} Leeds University Press, 57--62.

\bibitem{PG16} V.Patrangenaru, R. Guo and K. D. Yao (2016). {\em Nonparametric Inference for Location Parameters via Fr\'echet Functions},2016 Second International Symposium on Stochastic Models in Reliability Engineering, Life Science \& Operations Management, 254--262.

\bibitem{PaYagu:2016} V. Patrangenaru, K. D. Yao and R. Guo (2016). {\em Extrinsic means and antimeans}, Nonparametric Statistics, 161--178.

\bibitem{Su:2006} M. S. Sugathadasa (2006). {\em Affine and projective shape analysis with applications}, PhD dissertation, Texas Tech University.

\bibitem{Wa:1983} G. S. Watson (1983), {\em Statistics on Spheres}. University of Arkansas Lecture Notes in the Mathematical Sciences,  {\bf 6}.  Wiley-Interscience.


\end{thebibliography}
\end{document}